\documentstyle[12pt]{article}
\def\beq{\begin{equation}}
\def\eeq{\end{equation}}

\def\nabl{\nabla\!}

\def\thf{\baselineskip=\normalbaselineskip\multiply\baselineskip
by 5\divide\baselineskip by 4}
\thf

\def\nabl{\nabla\!}

\def\Ck{\kappa}
\def\ww{w}
\def\S{ {_{\rm S}} }
\def\xe{\xi}
\def\xo{\chi}
\def\PPi{H}
\def\Po{{\lambda}}

\def\rO{ {\ominus} }
\def\sO{ {\oslash} }
\def\rI{\, \grave{\,} }
\def\sI{ \acute{\,} }
\def\VV{ {\cal U} }

\begin{document}

\begin{center}
{\bf KALB-RAMOND COUPLED VORTEX FIBRATION MODEL 
FOR  RELATIVISTIC SUPERFLUID DYNAMICS}\\
\vskip 2cm
{\bf Brandon  Carter  and David Langlois} \\
\vskip 1cm
{\it D\'epartement d'Astrophysique Relativiste et de Cosmologie,\\
Centre National de la Recherche Scientifique, \\Observatoire de
Paris, 92195 Meudon, France}\\
\vskip 0.5cm
{\it and} \\
\vskip 0.5cm
{\it The Racah Institute of Physics, The Hebrew University,\\
Givat Ram, Jerusalem 91904, Israel} \\
\vskip 2cm
{1st June 1995}\\
\end{center} 

\vskip 1.6cm
 \noindent  
{\bf Abstract.} \it

The macroscopic dynamics of a rotating superfluid  deviates from that of a
simple perfect fluid due to the effect of vorticity quantisation, which
gives rise to a substructure of cosmic string type line defects that results 
in a local anisotropy whereby the effective average pressure in the direction 
of the  vortex lines is reduced below its value in lateral directions. 
Whereas previous descriptions of this effect have been restricted to a 
non-relativistic framework that is adequate for the treatment of liquid 
helium in a laboratory context, the present work provides a fully  
relativistic description of the kind required for application to rotating 
neutron star models. To start with, the general category of vortex fibration 
models needed for this purpose is set up on the basis of a Kalb-Ramond 
type variational principle. The appropriate specification of the particular 
model to be chosen within this category will ultimately be governed by the 
conclusions of microscopic investigations that have not yet been completed, 
but the results available so far suggest that a uniquely simple kind of model 
with an elegant dilatonic formulation should be tentatively adopted 
as a provisional choice so long as there is no indication that a more 
complicated alternative is needed.

\rm 

\vfill\eject

\section{Introduction.}
\medskip

Until now, calculations of the macroscopic effect of quantised vortices in
the superfluid layers of neutron stars have relied on the use of non
relativistic formalism whose most complete and detailed development is due
to Lindblom and Mendel$^{[1]}$. Despite the fact that relativistic
corrections are by no means small, their neglect has so far been physically
justifiable in the analysis of the most important observational effect for
which they are relevant, namely pulsar frequency glitches, due to the many
major uncertainties that are involved in the vortex pinning model that has
been proposed as an explanation$^{[2]}$. Although there is not yet any
urgent need for the higher accuracy that it can provide, the use of a fully
relativistic formulation is already wanted in order to satisfy a more
pressing requirement, namely that of compatibility with the fully
relativistic fluid models which have long been in standard use for
representing the the gross structure of neutron stars: the main drawback of
the mongrel combinations of relativistic and non relativistic elements
grafted together in the treatments that have been used so far is not so much
their lack of physical accuracy, but rather their lack of mathematical
coherence which makes them very unweildy to work with. The purpose of the
present work is to overcome this disadvantage by developing  a more
conveniently coherent approach in which the macroscopic effect of vortex
quantisation is allowed for in  a fully relativistic framework. The use of
this approach is exemplified here by the introduction of a simple prototype
model that is the natural relativistic generalisation of the simplest
special case within the category of non-relativistic models considered by
Lindblom and Mendel$^{[1]}$, allowing only for the bulk motion of the neutron
matter and ignoring the independent motion of the residual protons, whose
analysis is postponed for future work of a more detailed nature.

The concepts on which the present approach is based were originally developed 
not in neutron star theory but in a more exotic cosmological context, 
specifically that of axion field theory. It was Lund and Regge$^{[3]}$ who 
first drew attention to the analogy between  Kalb-Ramond coupled string motion 
in an axion field and vortex motion in an ordinary non relativistic 
superfluid. The immediate inspiration for the coherent relativistic
theory presented here was provided by the more recent analysis of Davis 
and Shellard$^{[4][5]}$, and its modification to allow for more general 
kinds of field by Ben Ya'acov$^{[6][7]}$. The outcome  effectively  
extends and completes the results of an earlier pionneering
investigation using a different but ultimately related approach
by Lebedev and Khalatnikov$^{[8]}$.

\vfill\eject
\section{The perfect fluid limit.}
\medskip

The proposed theory is obtained as the natural generalisation of an
appropriately reformulated version of ordinary ``barytropic" perfect fluid
theory given$^{[9]}$ by a Lagrangian variation principle which will now be
described.

By definition, perfect (relativistic) fluid medium is characterised by a
stress momentum energy density tensor $T^{\mu\nu}$ that is expressible in
terms of its timelike unit eigenvector $u^\mu$,  by the familiar formula 
$T^{\, \mu\nu}=\rho u^\mu u^\nu +P ( g^{\mu\nu}+u^\mu u^\nu)$, with   
$u^\mu u_\mu=-1$ where $g_{\mu\nu}$ denotes the background metric, and the 
scalars $\rho$ and $P$ are respectively the relativistic mass-energy density
and the pressure. The particularly simple case of an ordinary ``barytropic" 
perfect fluid model is specifiable by an ``equation of state" that 
determines $\rho$  as a function of the number density $n$ of a conserved 
current, or equivalently, in conjugate form,  that determines $P$ as a 
function of the relativistic chemical potential or ``effective mass" per 
particle variable $\mu$. The mutually conjugate functions for the mass 
density and the pressure are not independent but  are symmetrically related 
by a Legendre type transformation given by
$$\rho\{n\}+P\{\mu\}= n\mu\ , \hskip 1 cm \mu={d\rho\over dn}\ ,\hskip
1 cm n={dP\over d\mu}\ .\eqno(2.1)$$
These functions determine the corresponding sound speed, $c_\S$ say,
by a formula of the familiar form
$$c_\S^2={dP\over d\rho}={n\over \mu} {d\mu\over d n}\ .\eqno(2.2)$$

In such a model, as also on the generalisation to be presented in the
next section, a particularly important role is played by
 the {\it vorticity} 2-form $\ww_{\mu\nu}$ , which is defined as the exterior
derivative
$$ \ww_{\mu\nu} =2\nabl_{[\mu}\mu_{\nu]} , \eqno(2.3)$$
of the relevant 4-momentum covector, $\mu_\lambda$, which is the
dynamical conjugate of the conserved particle current vector
$n^\lambda$. The momentum covector is related  to its dual, the
corresponding momentum trivector, by
$$  \mu_\lambda={1\over 3!}\varepsilon_{\lambda\mu\nu\rho}
\PPi^{\mu\nu\rho}\ ,\eqno(2.4)$$
where $\varepsilon_{\lambda\mu\nu\rho}$ is the alternating tensor of the
4-dimensional spacetime background, while similarly the conserved current
vector is given in terms of its dual 3-form $N_{\mu\nu\rho}$ by
$$n^\lambda={1\over 3!}\varepsilon^{\lambda\mu\nu\rho} N_{\mu\nu\rho} \ .
\eqno(2.5)$$ 
The condition that the current is conserved, i.e. $\nabl_\mu n^\mu=0$,
 is expressible (in terms of
covariant derivation with respect to the  (flat or curved) spacetime metric
$g_{\mu\nu}$) as the closure of the corresponding 3-form, i.e. 
$$\nabl_{[\mu}N_{\nu\rho\sigma]}=0\  \eqno(2.6)$$
(where the square brackets denote index antisymmetrisation).
 
In the particular case of the barytropic perfect fluid model, though not in
the generalisation to be given in the following section, the momentum
covector and its dynamical conjugate the current vector will be expressible
in terms of the quantities introduced in (2.1) simply by
$$\mu_\lambda=\mu u_\lambda\ ,\hskip 1 cm n^\lambda=n  u^\lambda\ ,\eqno(2.7) $$
which implies
$$H^{\mu\nu\lambda}={\mu\over n}N^{\mu\nu\lambda}\ .\eqno(2.8)$$
In this case the complete set of equations of motion is given by
supplementing the kinematic relation (2.6) by a dynamical momentum transport
equation  which is obtainable from a  convective variational 
principle$^{[9]}$  in
the form
$$\PPi_{\mu\nu\rho}\nabl_\lambda\PPi^{\lambda\nu\rho}=0\ , \eqno(2.9)$$
whose more familar dual form is
$$ n^\mu\ww_{\mu\nu}=0\ . \eqno(2.10)$$

The most familiar category of equations of state to which this theory can be
applied is the ``polytropic" power law kind, $\rho\propto n^\gamma\ $, for a
constant ``polytropic index" $\gamma$. Particularly important examples are
the pressure free (``dust") case, $\gamma=1$ (for which the flow lines are
geodesics), and the ultrarelativistic (massless particle) gas case,
$\gamma=4/3$. In the generic case the variable $\mu$ will {\it not} vary
proportionally to $n$ so that the momentum 3-vector
$\PPi^{\mu\nu\rho}$ will {\it not} be simply identifiable with the index
raised version $N^{\mu\nu\rho}$ of the current 3-form $N_{\mu\nu\rho}$.
However such an identification {\it will} be possible in the special case of
the ``stiff" (linear) Zel'dovich equation of state$^{[10]}$, with
$\gamma=2$, as given by 
$$\rho={\hbar^3\over 12 m^2}N_{\mu\nu\rho}N^{\mu\nu\rho}\ , \hskip 0.8 cm
\PPi_{\mu\nu\rho}={\hbar^3\over m^2} N_{\mu\nu\rho}\ , \eqno(2.11)$$ 
where $m$ is a fixed mass scale (exact identification being obtainable by
choosing units such that $m^2 =\hbar^3$).

For an arbitrary equation of state there is a corresponding (zero temperature)
{\it superfluid} model$^{[11]}$ included as a particular application of the 
foregoing formalism when the (automatically conserved) vorticity is zero, 
since by (2.3) and (2.4) we shall then have 
$$\ww_{\mu\nu}=0\ \ \ \Rightarrow\ \ \ \nabl_\rho\PPi^{\rho\mu\nu}=0 \ ,
\eqno(2.12) $$ 
which is the necessary and locally sufficient integrability condition for the 
existence of a gauge scalar $\varphi$ that satisfies 
$$\mu_\nu=\nabl_\nu(\hbar\varphi)\ \ \Rightarrow\ \ \PPi^{\mu\nu\rho}
=\hbar\varepsilon^{\mu\nu\rho\sigma}\nabl_\sigma\varphi \ , \eqno(2.13)$$ 
and that represents a condensate order parameter which may be presumed to be
{\it axionic} in the sense of being periodic, so that it may have
string-like topological defects. The normalisation used here is such that
periodicity $2\pi$ is consistent with the usual quantisation condition that
the {\it conserved circulation} integral,
 $$\Ck=\oint\mu_\nu dx^\nu\ , \eqno(2.14)$$
round a defect should be a multiple of the standard Bohr unit $2\pi\hbar$.
In the special case of a Zel'dovich fluid the corresponding superfluid model
reduces exactly to the standard massless (Goldstone boson) axion field
model$^{[12][13][14]}$ with the coefficient $m$ representing the
corresponding Higgs mass scale.

\section{The vortex dynamical generalisation.}
\medskip

A Kalb-Ramond field is  definable, modulo addition of the exterior
derivative of an arbitrary 1-form $\alpha_\mu$, for {\it any} (perfect or 
other) {\it conserved} fluid as a gauge 2-form $B_{\mu\nu}$ from which the 
physical current 3-form is obtainable by exterior differentiation, i.e.
$$N_{\mu\nu\rho}=3\nabl_{[\mu}B_{\nu\rho]} \ , \hskip 0.8 cm
B_{\mu\nu}\cong B_{\mu\nu}+2\nabl_{[\mu}\alpha_{\nu]}\ ,\eqno(3.1)$$ 
the closure property (2.6) being  the Poincar\'e integrability condition that 
is necessary and locally sufficient for its existence. One can use it for 
setting up an appropriate action principle in which $B_{\mu\nu}$ is to be 
treated as a free field with the 2-form $\ww_{\mu\nu}$ as its source, which 
means that, instead of remaining a secondary construct obtained via (2.3), 
the vorticity must be promoted to the status of an independent -- albeit not 
entirely free -- field in its own right. This makes it eligible to contribute 
directly to the variational action in such a way as to allow at a macroscopic 
level for the effect of a smooth distribution of microscopic quantised 
vortices in a realistic (compressible) model for the treatment of superfluid 
dynamics on a large scale, in contexts such as that of a neutron star 
interior, as well as ordinary laboratory applications involving liquid 
$^4\!{\rm He}$, in the {\it zero temperature} limit. (To deal with the Landau 
type 2-constituent theory that is needed to allow for the independent entropy 
flux in a superfluid at finite temperature$^{[8][11][15]}$, a more 
intricate theory would be required.) 

The required form of the  Lagrangian  is given$^{[9]}$ by 
$${\cal L }=\Lambda-{1\over
4}\varepsilon^{\mu\nu\rho\sigma}B_{\mu\nu} \ww_{\rho\sigma}\ , \eqno(3.2)$$
where the ``master function", $\Lambda$, is a gauge independent scalar
function of the current 3-form $N_{\mu\nu\rho}$ (and also, of course of the
background space time metric), which is specified simply by taking
$\Lambda=-\rho$. The innovation to be considered here consists in allowing
the master function $\Lambda$ to depend not just on $N_{\mu\nu\rho}$ but
also on the vorticity 2-form $\ww_{\mu\nu}$. Consistently with the previous
definition$^{[9]}$ in the perfect fluid case, the {\it current momentum}
trivector $H^{\mu\nu\rho}$  and also, in the generic case, the relevant {\it
vorticity momentum} bivector $\Po^{\mu\nu}\ $, are specified as partial
derivatives  by the (fixed background) variation rule 
$$\delta\Lambda=-{1\over 3!}H^{\mu\nu\rho}\delta N_{\mu\nu\rho}-{1\over 2}
\Po^{\mu\nu}\delta\ww_{\mu\nu} \ .\eqno(3.3)$$ 
The final coupling term in (3.2) is contrived so that  requiring invariance of 
the corresponding action integral with respect to {\it free} variations of the 
gauge field $B_{\mu\nu}$, for a given value of $\ww_{\mu\nu}$, is equivalent 
to imposing a field equation 
$$\nabl_\mu \PPi^{\mu\nu\rho}={1\over 2}\varepsilon^{\nu\rho\mu\sigma} 
\ww_{\mu\sigma}\ .\eqno(3.4)$$ 
It can be seen  that this is precisely what is needed for the on shell value
of $\ww_{\mu\nu}$ to be consistent with the original definition by the
equations (2.3) and (2.4).

It remains to promulgate the rule governing the specification in (3.2) of the
source field $\ww_{\mu\nu}$, which obviously can not be entirely free since
otherwise the gauge invariance (3.1) would be violated and the system would in
consequence be overdetermined. To get the desired result, the vorticity is
introduced$^{[9]}$ via a fibration in a manner analogous to that by which 
the current was introduced in the convective variational formulation$^{[11]}$, 
but using a base space of only 2 (instead of 3) dimensions with a vorticity 
measure 2-form having components $\ww_{ij}$ with respect to local coordinates 
$\xo^i$ say for $i=1,2$, which act as a pair of freely variable 
(diffeomorphism gauge dependent) scalar fields. A projection $x^\mu\mapsto
\xo^{ i}$ structures the relevant spacetime neighbourhood as a fibre bundle 
over the material base manifold with  2 dimensional fibres interpretable as 
vorticity world sheets, and the base measure form $\ww_{ij}$ then determines 
the corresponding vorticity 2-form as its spacetime pullback: 
$$\ww_{\mu\nu}= \ww_{ij}\xo^{i}_{\, ,\mu} \xo^{j}_{\, ,\nu}\ .
\eqno(3.5)$$
It will always be possible locally to choose the coordinates to be flat
with respect to the symplectic structure $w_{ij}$ on the base space,
so that its non zero off diagonal components have unit magnitude, which
gives the explicit expression
$$w_{\mu\nu}= 2\chi^{_1}_{\, ,[\mu}\chi^{_2}_{\, ,\nu]}\ .\eqno(3.6)$$

The prescription (3.5) automatically ensures that $\ww_{\mu\nu}$ is closed,
$$\nabl_{[\mu}w_{\nu\rho]}=0\ , \eqno(3.7)$$
which is needed both as a condition for {\it gauge invariance} of the action
integral and also as an integrability condition for the field equation (3.4).
It also automatically ensures that $\ww_{\mu\nu}$ is algebraically degenerate, 
i.e. its determinant $\vert\ww\vert$ vanishes, so that it satisfies
$$\varepsilon^{\mu\nu\rho\sigma}
\ww_{\mu\nu} \ww_{\rho\sigma}=0\ , \hskip 0.8 cm  \ww_{\mu\nu}={1\over
2}\ww\, \varepsilon_{\mu\nu\rho\sigma} {\cal E}^{\rho\sigma}\ , \eqno(3.8)$$
where $\ww$ is the scalar vorticity magnitude and ${\cal E}^{\mu\nu}$ is the
unit bi-vector (as normalised by ${\cal E}^{\mu\nu}{\cal E}_{\nu\mu}=2$)
tangential to the congruence of vortex flux 2-surfaces characterised by 
constant values of the pair of comoving coordinates $\xo^i$. The allowed 
variations of $\ww_{\mu\nu}$ are those generated by infinitesimal displacements 
of the form $\delta\xo^i= -\xe^\mu\partial_\mu \xo^i $, for an arbitrary 
spacetime vector field $\xe^\mu$, so they are given by the corresponding Lie 
derivation formula which -- in view of (3.7) -- takes the form
$$\delta\ww_{\mu\nu}=2\nabl_{[\mu}\big(\ww_{\nu]\rho}\xe^\rho\big)
\ .\eqno(3.9)$$
Imposing that the action integral be invariant with respect to such
variations leads finally to a basic dynamical equation given by
$$\ww_{\mu\nu}(n^\nu-\nabl_\rho\Po^{\rho\nu})=0 \ , \eqno(3.10)$$
which is in  qualitative agreement with a form previously proposed by
Lebedev and Khalatnikov$^{[8]}$ on the basis of a different approach
whose relationship to the present analysis will be described in the
appendix.

The covariance of the foregoing formulation implies that the system will
satisfy a corresponding Noether identity of the standard form 
$$\nabl_\nu T^{\mu\nu}=0\eqno(3.11)$$ 
with 
$$T^{\mu\nu}=2{\partial{\cal L}\over\partial g_{\mu\nu}}+{\cal L}g^{\mu\nu}
 \ .\eqno(3.13)$$ 
Working out the extra term that will be induced on the right of (3.3) by a
virtual variation $\delta g_{\mu\nu}$ of the background metric, the
relevant stress energy momentum density tensor is found to be given by 
$$ T^\mu_{\ \nu}={1\over 2}H^{\mu\rho\sigma} N_{\nu\rho\sigma}+
\Po^{\mu\sigma}w_{\nu\sigma}+\Lambda g^\mu_{\,\nu}\ .\eqno(3.13)$$

\section{Conjugate reformulation.}
\medskip

It is useful for many purposes, and in particular for relating the present
approach to the previous work of Lebedev and Khalatnikov$^{[8]}$, to
reformulate the theory that has just been presented in a dynamically
conjugate version in terms of a modified master function $\Psi$ that is 
obtained from the original master function $\Lambda$ by a Legendre type 
transformation that is derivable as follows.

We start by replacing the Kalb Ramond form $B_{\mu\nu}$ by its dual
$$b^{\mu\nu}={1\over 2}\varepsilon^{\mu\nu\rho\sigma}
B_{\rho\sigma} \ , \eqno(4.1)$$
so that by (3.1) the current vector (2.5) is expressible simply by
$$n^\rho=\nabl_\nu b^{\rho\nu}\ ,\eqno(4.2)$$
which  entails the current conservation law
$$\nabl_\rho n^\rho=0\ ,\eqno(4.3)$$
as an immediate consequence.

Instead of postulating (3.1) or its dual version (4.2) as an axiom, we can
impose it via the variational principle by introducing an appropriate
Lagrange multiplier $\pi_\rho$ and replacing the original Lagrangian ${\cal
L}$ as given by (3.2) by the correspondingly augmented Lagrangian, whose
form, after subtraction of a divergence which does not affect the field
equations, is given by
$${\cal L}^\dagger={\cal L} +\pi_\rho\big(\nabl_\nu b^{\rho\nu}-n^\rho)
-\nabl_\nu\big(\pi_\rho b^{\rho\nu}\big) ,
\eqno(4.4)$$
in which $n^\rho$ and $\pi_\rho$ as well as $b^{\rho\sigma}$ are to be
considered as independent free variables, while as before $w_{\rho\sigma}$
is also to be considered as independent though not entirely free but given
by (3.5) which means that it is constrained to vary according to (3.9).
Invariance of the ensuing action with respect to variation of the current
$n^\mu$ itself gives the dynamical relation
$$\pi_\rho=\mu_\rho\eqno(4.5)$$
from which the requirement of invariance with respect to the other allowed
variations can be seen to give back the same dynamical equations as were
given by the original Lagrangian ${\cal L}$. 

Instead of considering the trivial relation (4.5) as a dynamical relation, 
let us now simply impose it as a defining restriction, so that the new 
Lagrangian (4.4) will reduce to the form
$${\cal L}^\dagger=\Psi+b^{\rho\sigma}\big(\nabl_\rho \mu_\sigma-{1\over2} 
w_{\rho\sigma}\big)\ ,\eqno(4.6) $$
where
$$\Psi=\Lambda-\mu_\rho n^\rho \ .\eqno(4.7)$$
It can be seen that the (fixed background)  variation of this function
 $\Psi$ will be simply expressible as
$$\delta\Psi=-n^\rho \delta\mu_\rho-{1\over 2}\Po^{\rho\sigma}\delta 
w_{\rho\sigma}\ ,\eqno(4.8)$$
which suggests that instead of regarding the components $n^\rho$ or
equivalently $N_{\mu\nu\rho}$ as  independent field variables from which
the components $\mu_\rho$ or equivalently $H^{\mu\nu\rho}$ are derived by
partial differentiation according to (3.3), it would be convenient to treat
the components $\mu_\rho$ as independent variables, from which the components 
$n^\rho$ are obtained by partial differentiation according to (4.8). 

This Legendre type transformation provides a new formulation in which the 
dual bivector $b^{\rho\sigma}$ to the Kalb Ramond form comes in as a Lagrange 
multiplier imposing the condition that the vorticity 2-form should be the 
exterior derivative of the momentum 1-form. In this conjugate reformulation, 
$\Psi$ is to be considered as a function just of the freely variable 
momentum covector $\mu_\rho$ and of the independent but constrained 
vorticity 2-form $w_{\rho\sigma}$ whose variation is governed by (3.9). Under 
these conditions the corresponding variation of the new Lagrangian (4.6) will 
be given by
$$\delta{\cal L }^\dagger-\nabl_\rho\big(b^{\rho\sigma}\delta\mu_\sigma\big)=
\big(\nabl_\sigma b^{\rho\sigma}-n^\rho\big)\delta \mu_\rho +
\big(\nabl_\rho\mu_\sigma-{1\over 2} w_{\rho\sigma}\big) \delta b^{\rho\sigma}
-{1\over 2}\big(b^{\rho\sigma}
+\Po^{\rho\sigma}\big) \delta w_{\rho\sigma}\ .\eqno(4.9)$$
As well as giving back the relation (4.2) and the specification (2.3) of the 
vorticity as the exterior derivative of the momentum, the use of (3.9) in 
the new variational principle gives back the main dynamical relation (3.10) in 
the form
$$j^\rho w_{\rho\sigma}=0\eqno(4.10)$$
where the modified current vector is given by the definition
$$j^\rho=\nabl_\sigma\big(b^{\rho\sigma}+\Po^{\rho\sigma}\big)=n^\rho
-\nabl_\sigma\lambda^{\sigma\rho}\ ,\eqno(4.11)$$
from which it is apparent that, like the particle current vector
$n^\mu$ itself, the modified current is also conserved:
$$\nabl_\rho j^\rho=0\ .\eqno(4.12)$$
Since the contribution to the action from the extra terms introduced in the
transformation (4.4) does not depend on the metric, it follows that the
stress momentum energy tensor derived from the modified Lagrangian ${\cal
L}^\dagger$, namely
$$T^{\rho}_{\ \sigma}=n^\rho\mu_\sigma-\Po^{\rho\nu}w_{\nu\sigma}+\Psi 
g^\rho_{\ \sigma}  \ ,\eqno(4.13)$$
will automatically be the same as the tensor (3.13) that was obtained from
the original Lagrangian ${\cal L}$.

\section{Explicit development.}
\medskip

The form of the master function $\Lambda$ for the original formulation given
in Section 3 can be made more explicit by expressing it in terms of the
three independent scalars that can be constructed from the tensors
$n^\rho$ and $w_{\rho\sigma}$ on which it depends. These can be
conveniently be taken to be the magnitude $n$ of the particle current vector
$n^\rho$ itself, the scalar magnitude $w$ of the vorticity covector
$w_{\rho\sigma}$ and the magnitude $\zeta$ of the associated Joukowsky lift
force density vector $\zeta_\rho$ as defined by
$$n^2=-n^\rho n_\rho\ , \hskip 1 cm w^2={1\over 2} 
w_{\rho\sigma}w^{\rho\sigma}\ , \hskip 1 cm \zeta^2=\zeta_\rho\zeta^\rho
\ , \eqno(5.1)$$
where the Joukowsky vector is defined as
$$\zeta_\rho=w_{\rho\sigma}n^\sigma={1\over 2} w{\cal E}^{\mu\nu}
N_{\mu\nu\rho}\ .\eqno(5.2)$$
This vector is interpretable as representing the volume density of force
that would be exerted on the vortices as an expression of the Magnus
effect, by the relative flux (if any) of the fluid according to the simple
formula originally derived by Joukowsky (or Zhukovski, depending on how one 
transliterates from Cyrillic) for flow past a long aerofoil. The
coefficients appearing in the derivation of $\Lambda$ can be expressed
explicitly in terms of partial derivatives with respect to these scalars in
the form
$${1\over 2} H^{\mu\nu\rho}=-{\partial \Lambda\over\partial n^2}
N^{\mu\nu\rho}-3{\partial \Lambda\over\partial\zeta^2}w{\cal E}^{[\mu\nu}
\zeta^{\rho]}\ , \eqno(5.3)$$
$$\lambda^{\rho\sigma}=-2{\partial\Lambda\over\partial w^2} w^{\rho\sigma}
-4{\partial\Lambda\over\partial\zeta^2} \zeta^{[\rho} n^{\sigma]}\ .
\eqno(5.4)$$
The stress energy momentum tensor can thereby be rewritten in the
manifestly symmetric form
$$T^\mu_{\ \nu}=-{\partial\Lambda\over\partial n^2}N^{\mu\sigma\tau}
N_{\nu\sigma\tau}-2{\partial\Lambda\over\partial w^2}w^{\mu\rho}
w_{\nu\rho}-2{\partial\Lambda\over\partial \zeta^2}\zeta^\mu \zeta_\nu
+\Big(\Lambda-2{\partial\Lambda\over\partial\zeta^2}\zeta^2\Big) 
g^\mu_{\ \nu} \ .\eqno(5.5)$$

An analogously explicit analysis of the conjugate formulation developed in 
Section 4 can be made by expressing the generalised pressure function $\Psi$
in terms of the three independent scalars that can be constructed from the
tensors $\mu_\rho$ and $w_{\rho\sigma}$ on which it depends. These can be
taken to be the magnitude $w$ of the vorticity vector, as already given by
(3.8) or (5.1) together with the magnitude $\mu$ of the momentum covector
and the magnitude $h$ of the helicity vector as given by
$$\mu^2=-\mu_\rho\mu^\rho\ , \hskip 1 cm h^2=h^\rho h_\rho\ ,\eqno(5.6)$$
where the helicity vector$^{[16]}$ is given by the definition 
$$ h^\mu={1\over 2}\varepsilon^{\mu\nu\rho\sigma}\mu_\nu
w_{\rho\sigma}=w\mu_\rho{\cal E}^{\rho\mu} \ , \eqno(5.7)$$
which, by the degeneracy property (3.6), is such as to ensure that it will
be conserved:
$$\nabl_\mu h^\mu=0\ .\eqno(5.8)$$
In terms of partial derivatives with respect to $w^2$, $\mu^2$ and $h^2$
one obtains the expressions
$$n^\rho=2{\partial\Psi\over\partial\mu^2}\mu^\rho+2{\partial\Psi\over
\partial h^2}w h_\sigma{\cal E}^{\sigma\rho} \ , \eqno(5.9)$$
$$\lambda^{\rho\sigma}=-2{\partial\Psi\over\partial w^2} w^{\rho\sigma}
-2{\partial\Psi\over\partial h^2} h_\nu H^{\nu\rho\sigma}\ .\eqno(5.10)$$
The corresponding manifestly symmetric expression for the stress momentum 
energy density tensor (4.13) is found to have the form
$$T^\rho_{\ \sigma}= 2{\partial\Psi\over\partial w^2} w^{\rho\tau}
w_{\tau\sigma}+ 2{\partial\Psi\over\partial\mu^2}\mu^\rho\mu_\sigma+
2{\partial\Psi\over\partial h^2}h^\rho h_\sigma+\Big(\Psi-2{\partial \Psi
\over\partial h^2} h^2\Big)g^\rho_{\ \sigma} \ .\eqno(5.11)$$

\section{Deviations from Perfect Fluid Limit. }
\medskip

Not only in the usual terrestrial laboratory experiments but also in the
neutron star applications for which the present theory is principally
intended, the relevant macroscopic angular velocities are sufficiently small
-- and the ensuing microscopic vortex lines therefore sufficiently widely
separated -- for it to be a good approximation to consider the effect of the
associated vorticity as a small perturbation to the dynamics. It will
therefore be useful to formulate the theory in terms of deviations from a
simple barytropic fluid model governed by an equation of state 
giving the zero vorticity density $\rho$ just as a function of the particle
number density $n$ or equivalently, in conjugate, form giving the zero
vorticity pressure $P$ as a function of the effective mass per particle
$\mu$. 

As a premilinary step, to prepare the way for a perturbation analysis
in which the deviations will be considered to be small, we shall start
without any loss of generality by considering deviations of unrestricted
amplitude, simply decomposing the master function $\Lambda$ as the sum
of the perfect fluid contribution, $\rho_{_\sO}$ say, to which it reduces when
the vorticity is zero, together with a vorticity dependent deviation
term $\acute\Lambda$ in the form
$$\Lambda=-\rho_{_\sO}+\Lambda\sI \ , \hskip 1 cm \rho_{_\sO}=\rho\{n\}
\ .\eqno(6.1)$$
We can also simply decompose the conjugate master function 
$\Psi$ in the analogous manner as the sum
of the perfect fluid contribution, $P_{_\rO}$ say, to which it reduces when
the vorticity is zero, together with a vorticity dependent deviation
term $\grave\Psi$ in the form
$$\Psi=P_{_\rO}+\Psi\rI \ , \hskip 1 cm P_{_\rO}=P\{\mu\}
\ .\eqno(6.2)$$
With respect to the vorticity tensor $w_{\rho\sigma}$ itself, the partial
derivatives of these deviation terms will be given by the same tensor
$\Po^{\rho\sigma}$ as for the total, but they will determine a reduced
momentum covector $\mu\sI\!_\rho$, and a reduced current vector
$n\!\rI^\rho$ according to the specifications
$$\delta\Lambda\sI=\mu\sI\!_\rho\delta n^\rho
-{1\over 2}\Po^{\rho\sigma}\delta w_{\rho\sigma} \ ,\eqno(6.3)$$
$$\delta\Psi\rI=-n\!\rI^\rho\delta \mu_\rho
-{1\over 2}\Po^{\rho\sigma}\delta w_{\rho\sigma} \ .\eqno(6.4)$$
Defining the particle current reference state unit vector $u_{_\sO}^\rho$
and the corresponding perfect fluid momentum contribution $\mu_{_\sO\rho}$
by
$$n^\rho=n u_{_\sO}^\rho\ ,\hskip 1 cm \mu_{_\sO\rho}=\mu_{_\sO}u_{_\sO\rho}
\ ,\hskip 1 cm \mu_{_\sO}=\mu\{n\}={d\rho_{_\sO}\over dn}\ ,\eqno(6.5)$$
one obtains the total momentum contribution from (6.1) in the form
$$\mu_\rho=\mu_{_\sO\rho}+\mu\sI\!_\rho\ .\eqno(6.6)$$
Similarly defining the momentum reference state unit vector $u_{_\rO}^\rho$
and the corresponding perfect fluid particle current contribution 
$n_{_\rO}^\rho$ by
$$\mu_\rho=\mu u_{_\rO\rho}\ ,\hskip 1 cm n_{_\rO}^\rho=n_{_\rO}u_{_\rO}^\rho
\ ,\hskip 1 cm n_{_\rO}=n\{\mu\}={dP_{_\rO}\over d\mu}\ ,\eqno(6.7)$$
one obtains the total particle current contribution from (6.2) in the form
$$n^\rho=n_{_\sO}^{\,\rho}+n\!\rI^\rho\ .\eqno(6.8)$$

Defining the particle current reference state perfect fluid contribution by
$$ T^{\rho\sigma}_{_\sO}=\rho_{_\sO} u_{_\sO}^\rho u_{_\sO}^\sigma +P_{_\sO}
 ( g^{\rho\sigma}+u_{_\sO}^\rho u_{_\sO}^\sigma) ,  
\hskip 1 cm P_{_\sO}=n\mu_{_\sO}-\rho_{_\sO}\ , \eqno(6.9)$$
one can express the total stress momentum energy density contribution in 
the form
$$ T^\rho_{\ \sigma}= T^\rho_{_\sO\sigma} +T_{\,}\sI^\rho_{\ \sigma}
\ , \eqno(6.10)$$
with
$$T_{\,}\sI^\rho_{\ \sigma}=n^\rho\mu\sI\!_\sigma+
\Po^{\rho\nu}w_{\sigma\nu}+\big(\Lambda\!\sI-n^\nu\mu\sI\!_\nu\big)
g^\rho_{\ \sigma}\ .\eqno(6.11)$$
This can be interpreted as the stress momentum energy density contribution
of a generalised kind of Stachel-Letelier type$^{[17][18][19]}$ ``string
fluid", given by the vorticity dependent action
contribution $\Lambda\sI$.

To obtain the analogous contribution from $\Psi\rI$ in the conjugate 
formulation we similarly define the momentum reference state perfect fluid
contribution by
$$ T^{\rho\sigma}_{_\rO}=\rho_{_\rO} u_{_\rO}^\rho u_{_\rO}^\sigma +P_{_\rO}
 ( g^{\rho\sigma}+u_{_\rO}^\rho u_{_\rO}^\sigma) ,  
\hskip 1 cm \rho_{_\rO}=n\mu_{_\rO}-P_{_\rO}\ . \eqno(6.12)$$
We can then express the total stress momentum energy density contribution in 
the alternative form
$$ T^\rho_{\ \sigma}= T^\rho_{_\rO\sigma} +T\rI^\rho_{\ \sigma}
\ , \eqno(6.13)$$
with
$$T\rI^\rho_{\ \sigma}=n\!\rI^\rho\mu_\sigma+
\Po^{\rho\nu}w_{\sigma\nu}+\Psi\rI
\, g^\rho_{\ \sigma}\ .\eqno(6.14)$$

It can be seen from (4.7) that the mutually conjugate deviation 
contributions will be related by
$$ \Lambda\sI-\Psi\rI = P_{_\rO}-P_{_\sO}+n^\nu\mu\sI\!_\nu
=\rho_{_\sO}-\rho_{_\rO}+n\!\rI^\nu\mu_\nu \ .\eqno(6.15)$$

In the limit as deviations from the zero vorticity state tend to zero,
it can be seen from the defining relations (2.1) that we shall have
$$ P_{_\rO}-P_{_\sO}\sim n(\mu-\mu_{_\sO})\ ,\hskip 1 cm
\rho_{_\sO}-\rho_{_\rO}\sim\mu(n-n_{_\rO})\ .\eqno(6.16)$$
Since by (6.6) and (6.8) we shall have
$$\mu_{_\sO}-\mu\sim u_{_\sO}^{\, \rho}\, \mu\sI\!_\rho\ , \hskip 1 cm
n_{_\rO}-n\sim u_{_\rO\rho}\, n\!\rI^\rho\ ,\eqno(6.17)$$
it can be seen from (6.15) that in this limit the deviations of the
two mutually conjugate kinds of master function will coincide, i.e.
we shall have
$$\Psi\rI\sim\Lambda\sI\ .\eqno(6.18)$$

\section{Asymptotically Separable Model for the Weak Vorticity Limit.}
\medskip

A further simplification can be obtained by postulating that the master 
function has a separable form such that the vorticity  dependence in the 
deviation term is contained in a negative factor, $-\Upsilon$ say, depending 
only on the scalar magnitude $w$, with a positive coefficient $\Phi^2$ that 
depends only on the relevant perfect fluid reference state: this state will 
be labelled by $_\sO$ when determined by the particle current magnitude $n$ 
so that the deviation $\Lambda\sI$ defined in (6.1) will have the form 
$\Lambda\sI =-\Phi^2_{_\sO}\Upsilon $;  alternatively the reference state
will be labelled by $_\rO$ when determined by the effective mass per particle 
$\mu$ so that the corresponding deviation defined by (6.2) will have the form
$\Psi\rI = -\Phi^2_{_\rO}\Upsilon $.
It must be noticed that, although in general a model that is separable in the 
particle number representation (6.1) will not be exactly separable in the 
corresponding chemical potential representation (6.2) and conversely, it can
 be seen from the work of the preceeding section that
the property of separability is nevertheless conserved by the Legendre 
transformation {\it in the weak vorticity limit}.  
Such an {\it asymptotically separable} form, as given by
$$\Lambda\sI \sim-\Phi^2_{_\sO}\Upsilon\{w\}\ , \hskip 1 cm
\Psi\rI \sim -\Phi^2_{_\rO} \Upsilon\{w\} \ ,\eqno(7.1)$$
in the  relevant weak vorticity limit characterised by $\Upsilon\rightarrow 
0$, is suggested by our recent analysis$^{[11]}$  of the average stress 
momentum energy density for an individual vortex cell. The results of this 
analysis can be matched in a very satisfactory manner by an appropriate 
choice of the single variable functions $\Upsilon$ and $\Phi$.

Dropping the label $_\sO$ or $_\rO$ in asymptotic formulae since there 
is no need to distinguish between the two kinds of reference state because
either interpretation would be valid, as illustrated by the relation 
$\Phi_{_\sO}\sim\Phi \sim \Phi_{_\rO}$, we can express the ensuing 
asymptotic forms of the particle current and momentum deviations given 
by (6.3) and (6.4) in the form
$$\mu\sI\!_\rho \sim {\Upsilon\over\mu} {d\Phi^2\over d n} \mu_\rho\ , 
\hskip  1 cm  n\!\rI^\rho \sim -{\Upsilon\over n}
 {d\Phi^2\over d\mu} n^\rho \ ,\eqno(7.2)$$
and we similarly obtain
$$\Po^{\mu\nu}\sim {1\over 2}\Po\,  \varepsilon^{\mu\nu\rho\sigma}
{\cal E}_{\rho\sigma}\ , \hskip 1 cm \Po\sim \Phi^2 
{d\Upsilon \over d w} \ . \eqno(7.3)$$
On substituting this in the expression (6.11) for the deviation of the stress 
momentum energy tensor from that of the particle current reference state, 
due to the vorticity dependent action contribution $\Lambda\sI$, we see that 
the asymptotic form of this deviation will be expressible in terms of the 
fundamental tangential and normal projection tensors associated with the 
flux 2-surfaces$^{[20][21]}$, as given respectively by
$$\eta^\rho_{\ \sigma}={\cal E}^\rho{_\nu}{\cal E}^\nu{_\sigma}\ ,
\eqno(7.4)$$
and
$$\perp^{\!\rho}_{\, \sigma}=g^\rho_{\ \sigma}-\eta^\rho_{\ \sigma}\ , 
\eqno(7.5)$$
in the form
$$T\sI^{\rho}_{\, \sigma}\sim n{d\Phi^2\over dn}\Upsilon u^\rho u_\sigma
-\Big(\Phi^2-n{d\Phi^2\over dn}\Big)\Upsilon \eta^\rho_{\ \sigma}
+{\mit\Pi}\sI \perp^{\!\rho}_{\, \sigma} \ , \eqno(7.6)$$
in which, relative to the particle current reference state, the relevant
effective vorticity pressure  ${\mit \Pi}\sI$, acting transversely to the
vortex flux lines, will be given by
$$ {\mit \Pi}\sI\sim w\Po-\Big(\Phi^2-n{d\Phi^2\over dn}\Big)
\Upsilon \ .\eqno(7.7) $$
In the equivalent conjugate formulation, the contribution from $\Psi\rI$, 
giving the deviation of the stress momentum energy density tensor with 
respect to that of the particle momentum reference state, will be
analogously expressible in the form
$$T\rI^{\rho}_{\, \sigma}\sim -\mu{d\Phi^2\over d\mu}\Upsilon u^\rho 
u_\sigma-\Phi^2 \Upsilon \eta^\rho_{\ \sigma}
+{\mit\Pi}\rI \perp^{\!\rho}_{\, \sigma} \ , \eqno(7.8)$$
in which, relative to the particle momentum reference state, the effective
vorticity pressure  ${\mit \Pi}\rI$, relative to the momentum reference
state, will be given by 
$$ {\mit \Pi}\rI\sim w\Po-\Phi^2 \Upsilon \ .\eqno(7.9) $$

In order to get the formulae (7.6), (7.7), (7.8) and (7.9) to agree
with the corresponding formulae as obtained in the preceeding work$^{[15]}$
by averaging over an individual vortex cell with circulation given by the 
standard quantum unit $\kappa=2\pi\hbar$, it suffices to take the
functions $\Phi$ and $\Upsilon$ to be given by
$$ \Phi^2={n\over\mu}\ , \hskip 1 cm \Upsilon={\hbar\over 4} w\,
{\rm ln}\Big\{ {w_{_\odot}\over w}\Big\} \ ,\eqno(7.10)$$
where $w_{_\odot} $ is a fixed cut off vorticity value -- whose exact
value is unimportant in the limit $w<<w_{_\odot}$ under consideration.
The specification (7.10) means that the function $\Phi$ is to be identified
with the dilation amplitude that has been shown to have a specially
important role in the newly developed variational formulation of perfect
fluid theory$^{[9]}$. The logarithmic derivative of this function
is expressible in terms of the corresponding sound speed (2.2) according
to either of the equivalent mutually conjugate specifications
$${n\over\Phi^2}{d\Phi^2\over d n}=1-c_\S^2\ , \hskip 1 cm
{\mu\over\Phi^2} {d\Phi^2\over d\mu}= c_\S^{-2}-1 \ .\eqno(7.11)$$
Since (7.10) also implies
$$\lambda w\sim \Phi^2\Upsilon \ ,\eqno(7.12)$$
the deviation with respect to the particle current reference state is
thus obtainable from (7.6) and (7.7) in the form
$$T\sI^\rho_{\ \sigma}\sim  \Phi^2\Upsilon\Big( (1-c_\S^2)u^\rho 
u_\sigma-c_\S^2  \eta^\rho_{\ \sigma}
+(1-c_\S^2) \perp^{\!\rho}_{\, \sigma}\Big) \ , \eqno(7.13)$$
while the conjugate formula for the deviation with respect to the particle
momentum reference state is similarly obtainable from (7.8) and
(7.9) in the form
$$T\rI^{\rho}_{\, \sigma}\sim \Phi^2\Upsilon\Big( (1-c_\S^{-2}) u^\rho 
u_\sigma-\eta^\rho_{\ \sigma}\Big)\ ,\hskip 1 cm
{\mit\Pi}\rI \sim 0 \ . \eqno(7.14)$$

The upshot of the foregoing analysis is that a master function given in 
terms of functions of the form (7.10) by
$$\Lambda=-\rho_{_\sO}-\Phi_{_\sO}^2 \Upsilon\{w\}\ 
+ o\{\Upsilon\},\eqno(7.15)$$
(where the label $_\sO$ indicates that the quantities concerned are to be
considered as functions just of the particle number density $n$)
provides a vortex fibration model whose stress momentum energy density
tensor is consistent with what has been derived$^{[15]}$ by averaging that
of an individual vortex cell in the weak vorticity limit.

So long as $w$ varies within a few orders of magnitude of some mean
value $\langle w \rangle$ that is itself a great many orders of magnitude
smaller than the the cut off value $w_{_\odot}$ appearing in (7.10), i.e.
$$w\approx \langle w \rangle\ ,\hskip 1 cm \langle w \rangle <<w_{_\odot}
\ , \eqno(7.16)$$
which will be a very good approximation in typical contexts that can be
envisaged for the application of the present theory, and in particular in 
the case of the neutron star matter for which it is principally intended, 
then the variation of the logarithmic factor in the formula (7.10) will be
negligible, so that it will suffice to replace it simply by a linear 
function, with a constant coefficient ${\cal K}$ proportional to the 
circulation $\kappa$ round an individual vortex, that will be given by
$$\Upsilon={\cal K} w \ , \hskip 1 cm {\cal K}={\kappa\over 8\pi} {\rm ln}
\Big\{ {w_{_\odot}\over \langle w\rangle}\Big\}=
{\hbar\over 4} {\rm ln}
\Big\{ {w_{_\odot}\over \langle w\rangle}\Big\} \ .\eqno(7.17)$$

\section{Dilatonic model.}
\medskip

As an ansatz for matching the available results obtained$^{[15]}$ from the
analysis of individual vortex cells, the asymptotic form (7.15) obtained in
the preceeding section is not the only possibility, but it is the the only
one that satisfies the simplifying condition that the master function
$\Lambda$ be dependent only on the first two scalars, $n$ and $w$, of the
triplet introduced in (5.1), or equivalently that the conjugate master
function $\Psi$ depend only on $w$ and the first scalar, $\mu$, of the pair
introduced in (5.6):
$${\partial \Lambda\over\partial\zeta}=0\ \hskip 0.6 cm\Leftrightarrow
\hskip 0.6 cm {\partial\Psi\over\partial h}=0 \ .\eqno(8.1)$$
Unlike an alternative simplifying assumption implicit in the earlier work of
Lebedev and Khalatnikov[8] which will be described in the appendix, the
condition (8.1) has the convenient feature of treating the mutually
conjugate functions $\Lambda$ and $\Psi$ on the same footing. The extent to
which such a simplifying condition is accurate in the weak vorticity limit
with which we are concerned remains to be checked by future work on
non-axisymmetric vortex cells, but until there is any evidence that a more
complicated ansatz may be needed, the most reasonable procedure is to adopt
what is obviously the simplest provisionally admissible supposition, namely
(8.1), as a tentative working hypothesis. 

Proceeding on this basis, which means that the general laws (3.3) and
(4.8) will simplify just to
$$\delta\Lambda=-\mu\, \delta n -\lambda\,\delta w\eqno(8.2)$$
and
$$\delta\Psi=n\,\delta\mu -\lambda\,\delta w\eqno(8.3) $$
with
$$\Psi=\Lambda+n\mu\ ,\eqno(8.4)$$
the relations (5.3), (5.4) or their conjugate forms (5.9), (5.10), will
reduce just to
$$\lambda^{\rho\sigma}={\lambda\over w} w^{\rho\sigma}\ ,
\eqno(8.5)$$
and
$$\mu^\rho=\Phi^{-2} n^\rho\ ,\hskip 1 cm \Phi^{-2}={\mu\over n}\ 
,\eqno(8.6)$$
so that the expression (4.13) for the stress momentum energy density tensor
will reduce simply to
$$T^{\rho}_{\ \sigma}=\Phi^{-2}n^\rho n_\sigma-\lambda w^{-1} 
w^{\rho\nu}w_{\nu\sigma}+\Psi g^\rho_{\ \sigma}  \ .\eqno(8.7)$$

As a compromise between the alternative mutually conjugate master functions
$\Lambda$ and $\Psi$, it is particularly convenient in such a case to work 
with a potential energy function $\VV$ defined by 
$$\VV=-{1\over 2}\big(\Lambda+\Psi\big)\ ,\eqno(8.8)$$ 
whose variation law, subject to (8.1), is obtainable from (8.2) and (8.3) in
the form
$$\delta \VV=\lambda\,\delta w+{\mu n\over \Phi}\,\delta\Phi 
 .\eqno(8.9)$$
This shows that -- except in the degenerate ``stiff" case for which $\mu$ is
just proportional to $n$ so that $\Phi$ is constant and $\delta\Phi$ vanishes
-- the potential $\VV$ can be used as a master function whose
specification in terms of the independent variables $w$ and $\Phi$
determines the original master function $\Lambda$ and its conjugate $\Psi$
by the formulae
$$\Lambda=-{n^2\over 2}\Phi^{-2} -\VV\ ,\hskip 1 cm
\Psi={\mu^2\over 2}\Phi^2 -\VV\, \eqno(8.10)$$
where the quantities $n^2$ and $\mu^2$  are obtained as functions of the
vorticity amplitude $w$ and the ``dilatonic" amplitude $\Phi$ via the
partial differential relations
$$\mu^2=\Phi^{-1}{\partial\VV\over \partial \Phi}\ ,\hskip 1 cm
n^2=\Phi^3{\partial \VV\over\partial \Phi}\ .\eqno(8.11)$$

Subject to the assumption (8.1) the potential defined by (8.9) can be used
to rewrite the Lagrangian (3.2) in the form
$${\cal L}=-{1\over 2}\Phi^{-2} n^2 -b^{\rho\sigma} 
\chi^{_1}_{\, ,\rho}\chi^{_2}_{\, ,\sigma}-\VV\{\Phi,w\}\ ,
\eqno(8.12) $$
which can be used as the basis of a new variational formulation in which the
independent field variables are classifiable in three subsets: the first 
consists of the dual Kalb Ramond bivector components $b^{\rho\sigma}$, as 
defined by (4.1) or (4.2), in terms of which the current amplitude $n$ is 
defined by
$$n^2=-\big(\nabl_\rho b_\nu^{\ \rho}\big)\nabl_\sigma b^{\nu\sigma}\ ;
\eqno(8.13)$$
the second subset consists of the vorticity base
coordinates $\chi^{_1}$ and $\chi^{_2}$ as defined by (3.5), in terms of
which, with respect to a flat symplectic gauge (3.6), the vorticity
amplitude will be given by
$$ w^2=2\chi^{_1}_{\, ,[\mu}\chi^{_2}_{\, ,\nu]}(\nabla^\mu\chi^{_1})
\nabla^\nu \chi^{_2} \ ;\eqno(8.14)$$
finally the third subset consists just of the dilatonic amplitude $\Phi$
which is to be considered here as an independently variable scalar in its
own right -- except in the degenerate ``stiff" case, for which $\VV$
depends only on $w$ with $\Phi$ acting merely as a fixed coupling constant.

An attractive feature of the ``dilatonic" formulation set up in this way
is that it allows the physically required asymptotic form (7.15) to be
matched by taking the new master function ${\cal U}$ to have a form that
is not just separable but more specifically linearly separable with 
respect to $\Phi^2$, meaning that it is expressible as
$$\VV\{\Phi, w\}=V\{\Phi\}+\Phi^2\Upsilon\{w\}\ ,\eqno(8.15)$$
for suitable single variable functions $V$ of $\Phi$ and $\Upsilon$ of
$w$. The condition of matching (7.15) in the weak vorticity limit 
leaves no freedom of choice in the specification of these single variable
functions: $\Upsilon$ here must be the same as the function 
that is denoted by the same symbol in  (7.15) and that is given explicitly
by (7.10) or with sufficient precision for practical purposes by (7.17),
while $V$ must be the same as the function denoted by the same symbol
in a recent discussion of the dilatonic formulation of the perfect fluid 
limit$^{[9]}$, which means that it is determined by the equation of state
for the mass density $\rho$ of the underlying perfect fluid as a function
of its conserved number density $n$ according to the parametric 
specification
$$V\{n\}=\rho\{n\}-{1\over 2}n{d\rho\{n\}\over d n}\ ,\hskip 1 cm
\Phi^{-2}=n^{-1}{d\rho\{n\}\over d n}. \eqno(8.16)$$
The exceptional Zel'dovich$^{[10]}$ case characterised by a fixed dilatonic
amplitude $\Phi=a$ say arises from  an equation of state of the
``stiff" type $\rho=n^2/2 a^2+b$ where $b$ is also fixed
(acting just as a cosmological constant) which gives $V=b$ and hence
$\VV=b+a^2\Upsilon$. 

In any such linearly separable model the vorticity coefficient $\lambda$ 
will be given just by
$$\lambda=\Phi^2{d\Upsilon\over dw}\ ,\eqno(8.17)$$
and, except in the  degenerate ``stiff" case (for which $dV/d\Phi$ is 
indeterminate), the original master function will have a value that is given
by a linearly separable expression of its own, namely
$$\Lambda=-{\Phi\over 2}{d V\over\ d\Phi}-V-2\Phi^2\Upsilon\ 
,\eqno(8.18)$$
while the conjugate function $\Psi$ turns out to have a corresponding
expression in which the dependence on $w$ drops out altogether, leaving
just
$$\Psi={\Phi\over 2}{d V\over d\Phi}-V \ .\eqno(8.19)$$

In applications for which the linear formula (7.17) for the vorticity
dependent factor $\Upsilon$ is considered to be sufficiently accurate,
so that (8.17) will take the specific form
$$\lambda={\cal K}\Phi^2\ ,\eqno(8.20)$$
while (8.11) will take the specific form
$$n^2=\Phi^3{dV\over d\Phi}+2{\cal K}\Phi^4 w\ ,\eqno(8.21)$$
the Lagrangian (8.12) of the dilatonic formulation can be written out
with a fully explicit presentation of the gradient dependence, which in 
this case will be homogeneously quadratic, as
$${\cal L}={1\over 2\Phi^2} \big(\nabl_\rho b_\nu^{\ \rho}\big)
\nabl_\sigma b^{\nu\sigma} -b^{\rho\sigma} 
\chi^{_1}_{\, ,\rho}\chi^{_2}_{\, ,\sigma}-{\cal K}\Phi^2\sqrt
{2\chi^{_1}_{\, ,[\mu}\chi^{_2}_{\, ,\nu]}(\nabla^\mu\chi^{_1})
\nabla^\nu \chi^{_2}} -V\{\Phi\}\ .\eqno(8.22) $$
In the application of the corresponding action principle,
variation with respect to the dual Kalb Ramond bivector components
$b^{\rho\sigma}$ gives the formula (2.3) for the vorticity form (3.6) in
terms of the momentum covector given by (8.6) in conjunction with the
definition (4.2); variation with respect to the vorticity base coordinates
$\chi^{_1}$ and $\chi^{_2}$ with the definition (8.5) gives the basic
dynamical equation of motion (3.10); while 
finally -- except in the ``stiff" case for which $\Phi$ is simply
fixed as a coupling constant -- variation with respect to
the dilatonic amplitude  itself gives  the equation (8.21) whose
solution (for a given form of the equation of state function $V\{\Phi\}$)
determines the value of this amplitude $\Phi$ as a function of the current
magnitude $n$ given by (8.13).

In  simple cases the correspondence (8.16) between the equation of state for
$\rho$ as a function of $n$ and the associated equation of state for
$V$ as a function of $\Phi$  can be made explicit as discussed in the
recent analysis of the perfect fluid case$^{[9]}$. The simplest example
of relevance as an approximation for neutron star matter is that
of the standard relativistic polytrope with index $\gamma=4/3$ as given,
for a constant coefficient $a$, by
$$\rho={3 a\over 2} n^{4/3}\hskip 0.6 cm \Rightarrow \hskip 0.6 cm
 V=2a^3 \Phi^4 \ .\eqno(8.23)$$

It is not to be expected that such a simple model could provide a completely
accurate description for the most general kinds of astrophysical and 
terrestrial applications. High accuracy would require an even more elaborate 
treatment allowing for the effect$^{[22]}$ of vortex lattice rigidity, and 
for many purposes in the context of neutron star applications it will also be
necessary to allow for the further complication of magnetic effects
involving the independently conducting proton fluid component$^{[1][2]}$.
However the most serious limitation of the above model may turn out to be
due to inadequate allowance for the effect of relative flow between the
averaged particle current and the vorticity surfaces. Such an effect was
deliberately excluded from consideration in the preceeding microscopic
analysis$^{[15]}$ of an individual vortex cell that provided the averages to
which the macroscopic model proposed here has been matched. A more difficult
microscopic analysis (lacking the cylindrical symmetry that facilitated the
previous calculation) still needs to be carried out by future work to
determine how the relevant averages are affected by the inclusion of the
effect of relative flow. Until the results of such a more complete
microscopic analysis are available, the simple model set up in the present
and preceeding section should be considered to be physically trustworthy
only in the limit of negligibly small values of the Joukowsky vector
(5.2) which can be considered as specifying the magnitude and direction of
the relative flow. 

We may sum up by saying that  within the subcategory characterised by the
ansatz (8.1) postulating that the master function $\Lambda$ depends only on 
the current magnitude $n$ and the vorticity scalar $w$ but not on the 
Joukowsky scalar $\zeta$, the condition of matching the results of our 
preliminary microscopic analysis$^{[15]}$ for the case $\zeta=0$ leads 
uniquely, as far as the weak vorticity limit is concerned, to a model of the 
kind set up in Section 7 and Section 8. A more complete microscopic analysis 
may show that within the general framework provided in Section 3 some other 
ansatz, such as the example$^{[8]}$ described in  Appendix B, whereby 
$\Lambda$ is made to depend also  on $\zeta$, will be necessary for matching 
cases for which $\zeta$ differs significantly from zero. Nevertheless the 
obvious ansatz of $\zeta$ independence has provided an elegant prototype 
model, which should provide a reasonably accurate description of stationary 
configurations in which there is no relative flow of the fluid relative to 
the vortex congruence, and perhaps even a moderately  realistic description 
under more general circumstances. Considered as a toy, this prototype model 
can also be used, as described in Appendix A, to provide a more general 
illustration of the  application of cosmic string theoretical concepts
to superfluid dynamics that was considered by Davis and Shellard$^{[4][5]}$ 
within the restrictive framework of the ``stiff" limit characterised by a 
fixed value of $\Phi$ with vanishing $V$. 

\bigskip
\bigskip\noindent
{\bf Acknowledgement.}
\medskip

The authors thank G.L. Comer, R.L.  Davis, I.M. Khalatnikov, 
L. Lindblom, X. Martin,  P. Peter, D. Priou, J. Shaham, and E.P.S. Shellard  
for instructive discussions.

\bigskip
\bigskip\noindent
{\bf Appendix A: The tie up with cosmic string theory.}
\medskip

The preceeding theory is designed to represent the macroscopically averaged
effect of a congruence of microscopic vortices of which each individual
member can be considered as an example of the kind of topological defect
known in a cosmological  context as a {\it global cosmic string}, of which
the simplest type is provided by the axion field theory in which the 
relevant underlying fluid model is of the ``stiff" kind in which the
dilatonic amplitude $\Phi$ has a constant value. (The qualification
``global" is needed here to distinguish such a relatively extended
configuration from the more strictly string-like gauge coupled case for
which, instead of being logarithmically divergent, the defect energy
distribution is locally confined: such ``local" string defects occur in the
familiar laboratory context of ordinary metallic superconductors, and are
also predicted -- in coexistence with the dynamically dominant global string
defects with which the present work is concerned -- in the interior of
neutron stars$^{[1][2]}$, where the relevant gauge coupled superconducting
current is constituted from protons.)

To see how the theory set up in Section 8 is to be interpreted in the terms
of the technical machinery originally developed to describe string defects
of the cosmic type, it is to be remarked that the general form (8.7) that is
obtained from (8.1) for the stress momentum energy density tensor is
decomposable in a natural manner as the sum,
$$T^\rho_{\ \sigma}=T^\rho_{\circ\sigma}-{w\over\kappa}T
\eta^\rho_{\ \sigma}\ ,\eqno(A1)$$
of a spacially isotropic (non-barytropic) perfect fluid type contribution
$$T^\rho_{\circ\sigma}=n^\rho \mu_\sigma+
\big(\Psi +\lambda w\big) g^\rho_{\ \sigma} \eqno(A2) $$
together with a term of the form that would be given by a
Stachel-Letelier$^{[17][18][19]}$ type ``string fluid", whereby each
individual vortex cell contributes as if it were a string of the simple 
degenerate (longitudinally Lorentz invariant) Goto-Nambu type with
fundamental tensor $\eta^\rho_{\sigma}$ and tension $T$ given by
$$T=\kappa\lambda ,\eqno(A3)$$
since the vorticity $w$ is interpretable as due to a flux of $w/\kappa$
distinct microscopic vortices per unit area, where $\kappa$ denotes the
constant momentum circulation associated with each individual vortex.

More particularly, for the  specific model of the linearly separable kind
characterised by (8.15) with the vorticity dependent factor $\Upsilon$ given
by the linear  formula (7.17) on which (8.22) is based, the tension of each
such (degenerate Goto-Nambu type) string will be given by
$$T=\kappa{\cal K}\Phi^2={\pi\hbar^2\over 2}\Phi^2
{\rm ln}\Big\{ {w_{_\odot}\over \langle w\rangle}\Big\} \ .\eqno(A4)$$
and in this case the perfect fluid contribution (A2) will be expressible
simply by
$$T^\rho_{\circ\sigma}=n^\rho \mu_\sigma+
\big( {1\over 2}n\mu-V\big) g^\rho_{\ \sigma} \eqno(A5) $$
which has just the form that would be obtained from (8.22) if the
vorticity coupling coefficient ${\cal K}$ in it were set to zero.

Since the closure condition (3.7) gives the kinematic identity
$$\nabl_\nu\big(w{\cal E}^{\nu\rho}\big)=0\hskip 0.6 cm\Rightarrow
\hskip 0.6 cm w_{\rho\sigma}\nabl_\nu w^{\nu\sigma}=w\perp^{\!\nu}_{\,\rho}
\nabl_\nu w-w^2 K_\rho\ ,\eqno(A6)$$
where $K_\rho$ is the geometric curvature vector of the vortex sheet,
as given$^{[21]}$ in terms of its fundamental tensor (7.6) by
$$ K_\rho=\eta^\nu_{\ \sigma}\nabl_\nu \eta^\sigma_{\ \rho}=
{\cal E}^\nu_{\ \sigma}\nabl_\nu{\cal E}^\sigma_{\ \rho}\ , \eqno(A7)$$
the proportionality relation (8.5) can be
used to reduce the dynamical equation of motion (3.10) to the form
$$\zeta_\rho=w\big(\perp^{\!\nu}_{\, \rho}\nabl_\nu\lambda-\lambda\, K_\rho
\big)\ \eqno(A8)$$ 
for any model of the subcategory characterised by (8.1). Since the vorticity
$w$ is interpretable as due to a flux of $w/\kappa$ distinct string-like
vortices per unit area, the Joukowski force density $\zeta_\rho$, as defined
by (5.2), can be interpreted as representing a corresponding average
Joukowski force per unit length, $Z_\rho$ say, acting on each individual
vortex cell, that will be given by
$$Z_\rho={\kappa\over w}\zeta_\rho={1\over 2}\kappa {\cal E}^{\mu\nu}
N_{\mu\nu\rho} \ .\eqno(A9)$$
The equation of motion (A8) can thus be seen to be expressible as the 
condition that this Joukowski force per unit length should be 
given in terms of the corresponding effective vortex tension (A3) 
simply by
$$Z_\rho+ T K_\rho=\perp^{\!\nu}_{\, \rho}\nabl_\rho T \ .\eqno(A10)$$
This can be recognised as a generalisation of the standard equation of
motion for a global cosmic string in an axion field 
background$^{[13][14]}$, to which it reduces when the tension $T$ is
constant so that the gradient term on the right drops out. The constant 
tension case arises for the model (8.22) obtained from (7.17) to which
(A4) applies when the equation of state is of the ``stiff" type for
which $\Phi$ is constant.

\bigskip
\bigskip\noindent
{\bf Appendix B: The Lebedev Khalatnikov subcategory.}
\medskip

It is easy to confirm that the category of theoretical models obtained by
the convective variational approach described in Section 3 includes as a
subcategory the class of models originally set up by Lebedev and
Khalatnikov$^{[8]}$. To show this it suffices to identify their generalised
pressure function $\Psi$ with the Legendre transformed master function
$\Psi$  on which the reformulated variational principle of the Section 4 is
based, since on this basis our equation of motion (4.10) agrees precisely
with the one that they obtained using a Clebsch type variational procedure.
Instead of using a Lagrangian of the form (4.6) in which the master function
is supplemented by the appropriate Kalb Ramond type coupling term, the
variational formulation proposed by Lebedev and Khalatnikov used $\Psi$ by
itself as  Lagrangian, the cost of this apparent simplification being the
need to consider the relevant momentum covector $\mu_\rho$ not as an
independent field variable in its own right but merely as a derived
quantity, specified by an appropriate set of Clebsch type potentials and
their gradients. The question of whether one prefers to work with a set of
scalar Clebsch type potentials as in the approach pioneered by Lebedev and
Khalatnikov, or with a single antisymmetric tensorial potential of Kalb
Ramond type as in the approach used here is merely a matter of taste and
convenience in view of the ultimate agreement of the ensuing field
equations. The Clebsch formulation has the technical advantage of using
fewer independent field components but has the drawback that a larger number
of different alphabetical letters are needed to characterise them. From the
point of view of mathematical elegance and ease of formal manipulation, the
Kalb Ramond formulation (either in its original version as given in the
 section 3 or the dual version of section 4) is more
satisfactory: it involves a greater degree of gauge dependence but has the
advantage of  being more economical in its use of algebraic symbols.

Although the postulates of the Lebedev Khalatnikov theory are thus confirmed
to be consistent with those adopted here, their category of models was
restricted to a subclass of those considered in Section 4  by the
supposition that the master function $\Psi$ depended only on $\mu$ and $h$
but not on the third independent scalar $w$, i.e. they assumed that the form
of the  master function would be characterised by 
$${\partial \Psi\over\partial w^2}=0\ .\eqno(B1)$$ 
Subject to this restriction the variation of $\Psi$ would be given simply by
$$\delta\Psi=\ell_\rho\delta h^\rho-m^\rho\delta\mu_\rho\ ,\eqno(B2)$$
with
$$\ell_\rho=2{\partial\Psi\over\partial h^2} h_\rho\ ,\hskip 1 cm
m^\rho=2{\partial\Psi\over\partial\mu^2}\mu^\rho \ .\eqno(B3)$$
In terms of these quantities, the stress momentum energy density tensor 
(5.11) will be reducible to the simple form
$$T^\rho_{\ \sigma}= m^\rho\mu_\sigma+h^\rho\ell_\sigma+\big(\Psi-
\ell_\nu h^\nu\big) g^\rho_{\ \sigma}\ ,\eqno(B4)$$
which agrees with the expression given by Lebedev and Khalatnikov$^{[8]}$.
However a small discrepancy does occur in their expression for the 
modified current vector (4.11), which omitted the final term of the 
formula
$$j^\rho=m^\rho-H^{\rho\sigma\nu}\nabl_\sigma \ell_\nu
+2w\ell_\nu{\cal E}^{\nu\rho} \eqno(B5)$$
that we obtain in this case.

The choice of the particular kind of model introduced in Section 7 and
Section 8 to match the results of our recent microscopic analysis$^{[15]}$
for the case $\zeta=0$  was determined by the obvious ansatz that the master
function $\Lambda$ should not depend on the Joukowsky scalar $\zeta$, but 
only on $n$ and $w$.  We saw that in the weak vorticity limit under
consideration, this ansatz was equivalent to the postulate that the
conjugate master function $\Psi$ should depend only on $\mu$ and $w$ but not
on the helicity scalar $h$. However there is not any obstacle to matching
the same results$^{[15]}$ by a model characterised by the alternative
Lebedev - Khalatnikov ansatz to the effect that $\Psi$ should instead depend
on $\mu$ and $h$ but not on $w$. This can be done simply by replacing $w$ in
the deviation term given by (7.1) by the ratio $h/\mu$ which is
the same as $w$ when $\zeta$ vanishes. It can be checked directly that the model
specified by the deviation term
$$\Psi\rI\sim - {\cal K}\Phi_{_\rO}^2{h\over\mu}  \eqno(B6)$$
that is obtained in this way will have a current vector and a stress
momentum energy density tensor that agree with those obtained from the
ansatz adopted in Section 7 for small $w$ provided $\zeta$ vanishes. 
However this ansatz leads to a more complicated form for the conjugate 
formulation which can be seen to have an explicit $\zeta$ dependence given by
$$\Lambda\sI\sim -{\cal K}\Phi_{_\sO}^2\sqrt{w^2-{\zeta^2\over n^2} }
\ .    \eqno(B7)$$
Until further evidence is provided by a more complete microscopic 
analysis, the criterion of simplicity would seem to rule against this
latter model in favour of the separable model proposed in Section 7.
However it may well turn out that neither alternative is adequate for
cases in which $\zeta$ is too large to be neglected.


\bigskip


\begin{thebibliography}{99}

\bibitem{[1]}  G. Mendel, L. Lindblom, {\it Ann. Phys.} {\bf 205}, 110 (1991).
\smallskip
\bibitem{[2]}  J.A. Sauls, in {\it Timing Neutron Stars}, ed H. Ogelman, E.P.J.
Van Heuvel, {\it pp} 457 - 489 (Kluwer, Dordrecht, 1988).
\smallskip
\bibitem{[3]}  F. Lunde, T. Regge, {\it Phys. Rev.} {\bf D14}, 1524 (1976).
\smallskip
\bibitem{[4]}  R.L. Davis, E.P.S. Shellard, {\it Phys. Lett.} {\bf B214,} 219 (1988).
\smallskip
\bibitem{[5]}  R.L. Davis, E.P.S. Shellard, {\it Phys. Rev. Lett.} {\bf 63},
2029 (1989).
\smallskip
\bibitem{[6]} U. Ben-Ya'acov, {\it Phys. Rev.} {\bf D44}, 2452 (1991).
\smallskip
\bibitem{[7]} U. Ben-Ya'acov, {\it Nucl. Phys.} {\bf B382}, 592 (1992). 
\smallskip
\bibitem{[8]} V.V. Lebedev, I.M. Khalatnikov, {\it Sov. Phys. J.E.T.P.} {\bf 56},
923 (1982).
\smallskip
\bibitem{[9]} B. Carter, {\it Class. Quantum. Grav.} {\bf 11}, 2013 (1994).
\smallskip
\bibitem{[10]}  Ya. B. Zel'dovich, {\it Sov. Phys. J.E.T.P.} {\bf 41}, 1143 (1962).
\smallskip
\bibitem{[11]}  B. Carter, I.M. Khalatnikov, {\it Phys. Rev.} {\bf D45 }, 4536 (1992).
\smallskip
\bibitem{[12]}  E. Witten, {\it Phys. Lett.} {\bf B153}, 243 (1985).
\smallskip
\bibitem{[13]}  A. Vilenkin, T. Vachaspati, {\it Phys. Rev.}  {\bf D35}, 1138 (1987).
\smallskip
\bibitem{[14]}  M. Sakellariadou,  {\it Phys. Rev.}  {\bf D44}, 3767 (1991).
\smallskip
\bibitem{[15]}  B. Carter, D. Langlois {\it Phys. Rev.} {\bf D}, sous presse (1995).
\smallskip
\bibitem{[16]} B. Carter, I.M. Khalatnikov, {\it Ann. Phys} {\bf 219}, 243 (1992).
\smallskip
\bibitem{[17]} J. Stachel, {\it Phys. Rev.} {\bf D21}, 2171 (1980).
\smallskip
\bibitem{[18]} P.S. Letelier, {\it Phys. Rev.} {\bf D20}, 1294 (1979).
\smallskip
\bibitem{[19]} W. Kopczynski, {\it Phys. Rev.} {\bf D36}, 3582 (1987).
\smallskip
\bibitem{[20]}  B. Carter, {\it Phys. Lett.} {\bf B228}, 446 (1989).
\smallskip
\bibitem{[21]}  B. Carter, {\it Class. Quantum. Grav.}  {\bf 9}, S19 (1992).
\smallskip
\bibitem{[22]} A.F. Andreev, M.Yu. Kagan, {\it Sov. Phys. J.E.T.P.} {\bf 59}, 318 
(1982).
\end{thebibliography}
\end{document}